\documentclass[11pt]{article}
\usepackage[utf8]{inputenc}
\usepackage[T1]{fontenc}
\usepackage{lmodern}
\usepackage[margin=1in]{geometry}
\usepackage{amsmath,amssymb}
\usepackage{graphicx}
\usepackage{float}
\usepackage{booktabs}
\usepackage{hyperref}
\usepackage{xcolor}
\usepackage{natbib}
\usepackage{enumitem}
\usepackage{placeins}
\usepackage{xurl}
\usepackage{microtype}
\usepackage{listings}
\usepackage{multirow}
\hypersetup{colorlinks=true, linkcolor=blue!70!black, citecolor=blue!70!black,
            urlcolor=blue!70!black}

\newcommand{\walk}{\textsc{walk}}
\newcommand{\trail}{\textsc{trail}}
\newcommand{\acyc}{\textsc{acyclic}}
\newcommand{\simplem}{\textsc{simple}}

\title{Same Pattern, Different Answer:\\
       A Reference Semantics and Divergence Map for\\
       GQL and SQL/PGQ Path Patterns}
\author{Madhulatha Mandarapu\thanks{madhulatha@samyama.ai}
        \and Sandeep Kunkunuru\thanks{sandeep@samyama.ai}}
\date{VaidhyaMegha Private Limited, India\\[2pt]\url{https://samyama.ai/}\\[8pt]
      September 2026}

\begin{document}
\maketitle

\begin{abstract}
GQL (ISO/IEC 39075:2024) is the first international standard for a graph query
language, and SQL/PGQ (ISO/IEC 9075-16:2023) embeds the same pattern-matching core,
GPML, in SQL. Both fix a precise semantics for path patterns: four path modes, four
selectors, quantified segments. What engines compute for those patterns has never been
measured. Work on GQL is theoretical and runs no engine; cross-engine work measures
performance and normalises the semantics away.
We build an executable reference semantics for the GPML path core and gate it against
the six worked answers printed in the standard's own reference exposition. We then run
a 17-construct suite against six releases of five products --- K\`uzu, DuckPGQ, Neo4j at 5.26 and 2026.04,
Memgraph, Apache AGE --- scoring each cell as conforming, diverging, rejected, or
inexpressible in that dialect.
Nine of seventeen constructs draw more than one answer across the engines that
accepted them. Of 26 disagreements, 15 are silent --- the query runs, returns a
different multiset, raises no error --- a silence ratio of 0.58. Scoring each divergence
against the path mode each engine's manual declares separates language from
implementation: 9 of 15 are documented differences, 6 are departures from the standard
the engine implements. A metamorphic layer needing no
reference semantics finds 39 self-consistency violations, yielding two minimal defects.
No engine we measured parses the order the standard's own examples are written in,
except the authors' own, which added it after this suite reported the gap. Suite, semantics,
reproducers and results are open source.
\end{abstract}

\section{Introduction}

For thirty years, querying property graphs meant querying a dialect. Cypher, PGQL,
GSQL and Gremlin each drew a different line around what a path is and which paths a
pattern returns~\citep{angles2017foundations}. In 2023 and 2024 that changed twice:
SQL:2023 gained Part 16, SQL/PGQ~\citep{iso9075_16}, and GQL was
ratified as ISO/IEC 39075~\citep{iso39075}, the first international standard for a
graph query language. Both build on the same pattern-matching core, GPML, whose
reference exposition was written by members of the standards
committee~\citep{deutsch2022gpml}.

GPML makes an unusually explicit semantic commitment. A path pattern may carry a
\emph{restrictor} --- \trail{} (no repeated edges), \acyc{} (no repeated nodes),
\simplem{} (no repeated nodes except that the first and last may coincide) --- and a
\emph{selector} --- \textsc{all}, \textsc{any}, \textsc{all shortest},
\textsc{any shortest}. With no restrictor the matched object is an unrestricted path,
which graph theory calls a walk. Selectors are applied after restrictors. Every
unbounded quantifier must sit inside the scope of one or the other, so that the answer
is finite; a \emph{bounded} quantifier needs neither.

So the standard answers, unambiguously, what \verb|(x)-[e]->{1,3}(y)| returns. The
question this paper asks is what engines return.

Nobody has measured it. The 2023--2026 literature on GQL is uniformly theoretical: a
formal semantics~\citep{francis2023digest}, expressive-power
separations~\citep{gheerbrant2025expressive}, and a mechanised small-step semantics
with a soundness proof~\citep{thimmaiah2026mgql}. None of them runs a database. The
one recent cross-engine paper that touches path semantics is about
performance~\citep{rivera2026path}: it observes in passing that Neo4j and Memgraph
enforce trail during matching while K\`uzu and DuckDB apply it afterwards, but all of
its queries use trail semantics, so the answers are normalised and no divergence is
measured. Vendor documentation offers pairwise prose lists, self-reported and not
executable.

\paragraph{Contributions.}
\begin{enumerate}[leftmargin=1.4em,itemsep=2pt]
\item \textbf{An executable reference semantics} for the GPML path core
      (Section~\ref{sec:ref}). It consults no engine, which is the point: differential
      testing across engines can establish that two systems disagree but not which is
      right. It is gated against the six worked answers the standard's reference
      exposition prints (Section~\ref{sec:gateb}).
\item \textbf{A 17-construct conformance suite} and the first answer-level divergence
      map over six releases of five products (Sections~\ref{sec:suite}--\ref{sec:map}).
\item \textbf{The silence ratio}, a statistic no prior work reports: the share of
      disagreements the user is not told about (Section~\ref{sec:map}).
\item \textbf{An attribution axis} separating a divergence that the engine's own
      documentation explains from one that does not (Section~\ref{sec:attribution}).
\item \textbf{A reference-free metamorphic layer} and the two engine defects it finds,
      each with a standalone reproducer (Section~\ref{sec:meta}).
\end{enumerate}

\section{Background: the GPML path core}
\label{sec:background}

A property graph is $(N, E, \rho, \lambda, \pi)$: disjoint node and edge identifier
sets, a map $\rho$ sending an edge to an ordered pair (directed) or an unordered pair
(undirected), a map $\lambda$ from elements to \emph{sets} of labels, and a partial map
$\pi$ from an element and a property name to a value. Two distinct edges may connect
the same pair of nodes, and self-loops are allowed~\citep{deutsch2022gpml}.

A \emph{path} is an alternating sequence $n_0, e_1, n_1, \dots, e_k, n_k$ that starts
and ends with a node, consecutive nodes being connected by the edge between them.
Traversal orientation is not part of the sequence. This is what graph theory calls a
walk; the standard calls it a path, and we follow the standard.

\begin{table}[H]
\centering
\begin{tabular}{@{}lll@{}}
\toprule
& Keyword & Condition \\
\midrule
Restrictors
& \trail{}   & no repeated edges \\
& \acyc{}    & no repeated nodes \\
& \simplem{} & no repeated nodes, except that the first and last may be the same \\
& (none)     & unrestricted --- a walk \\
\addlinespace
Selectors
& \textsc{all}           & every match \\
& \textsc{any}           & one match per endpoint pair, unspecified which \\
& \textsc{all shortest}  & every match of minimum length, per endpoint pair \\
& \textsc{any shortest}  & one match of minimum length per endpoint pair \\
\bottomrule
\end{tabular}
\caption{The path restrictors and path selectors of GPML. Selectors partition the solution space
on the endpoints and select within each partition; when both appear, the selector is
applied after the restrictor.}
\label{tab:modes}
\end{table}

Two consequences matter for what follows and are easy to get wrong.

\paragraph{A bounded quantifier with no restrictor means walk.} The standard's own
worked example is
\begin{center}
{\footnotesize\verb|(p WHERE p.owner='Natalia')->{1,10}(q WHERE q.owner='Mike')->{1,10}(r WHERE r.owner='Scott')|}
\end{center}
for which it states that a particular solution
traverses one edge twice, and adds that this solution ``fails the restrictor \trail{}
(as well as \simplem{} and \acyc{})''. A bounded quantifier is finite on its own, so no
restrictor is implied.

\paragraph{The quantifier binds the edge pattern, not the node pattern.} In the same
example the printed solution passes \emph{through} a node that does not satisfy the
segment's node pattern on its way to one that does. Interior nodes under a quantifier
are unconstrained; the node pattern binds at the end of the repetition.

\section{An executable reference semantics}
\label{sec:ref}

We implement the definitions of Section~\ref{sec:background} directly, with no
optimisation and no engine in the loop. A pattern is a start node pattern followed by
quantified segments; evaluation enumerates paths, filters by the restrictor, then
applies the selector, in that order.

Three design points are worth stating because they are where a naive oracle goes
wrong.

\paragraph{Unbounded quantifiers get a sound finite bound, derived per case.}
Under \trail{} no edge repeats, so at most $|E|$ edges; under \acyc{} or \simplem{} at
most $|N|$; under an unrestricted walk with a shortest selector, a shortest walk
between two nodes is a simple path, so at most $|N|-1$. An unbounded quantifier under
an unrestricted walk with \textsc{all} is forbidden by the standard, and with
\textsc{any} its admissible set is infinite, so no finite oracle exists; the
implementation raises rather than guesses.

\paragraph{Nondeterminism is specified, not divergent.} \textsc{any} and
\textsc{any shortest} fix how many paths are returned per endpoint partition, not which
ones. The oracle therefore returns a \emph{specification} --- an exact multiset when the
selector is deterministic, and otherwise an admissible set together with a
per-partition count. An engine is judged against that specification. We never score a
permitted choice as a divergence.

\paragraph{Prefix pruning must respect the restrictor's shape.} \trail{} and \acyc{}
are prefix-closed, so a prefix that already repeats an edge or a node can be pruned.
\simplem{} is not: a prefix whose first and last nodes coincide is legal only if the
path stops there, so pruning uses the weaker interior condition.

\subsection{Gate B: reproducing the standard's published answers}
\label{sec:gateb}

The reference exposition prints the exact path bindings six of its worked queries
return over the property graph of its Figure~1. We require our implementation to
reproduce all six before any engine is measured. The graph is a figure rather than
text, so its \verb|Transfer| sub-graph is reconstructed from the paths the paper itself
prints; every edge is witnessed by at least one printed path, and the reconstruction is
not asserted but tested --- a wrong edge set would break at least one of the six.

The gate has teeth. An oracle that applies the segment's node pattern at every
repetition, rather than only at the end of it, fails three of the six checks --- and
that is a reading of Section~\ref{sec:background} a careful person can arrive at, not a
typo. An error of that shape would corrupt every cell of the map without producing a
single symptom, because the map has no other yardstick to disagree with.

\section{The suite}
\label{sec:suite}

Seventeen cases, each a construct where the standard makes a commitment and an
implementation has freedom. The engines are K\`uzu~\citep{feng2023kuzu}, DuckPGQ~\citep{wolde2023duckpgq},
Neo4j, Memgraph and Apache AGE. Each case carries the abstract pattern, the natural
rendering in openCypher, and the natural rendering in SQL/PGQ. Where a dialect has no
way to express the construct, the entry is empty and the cell scores
\textsc{inexpressible} --- absence of syntax is a fact about the language, not a defect.

Graphs are deliberately tiny, so that every expected answer is checkable by hand and
every divergence is a one-screen reproducer. Six fixtures suffice: a directed
two-cycle; a graph with parallel edges and a three-cycle; a self-loop with a two-cycle;
a single edge; a two-hop chain whose interior node lacks the endpoints' label; and the
reconstructed \verb|Transfer| sub-graph of the standard's Figure~1.

Comparison is at the level of the multiset of endpoint identifier pairs. Every engine
can produce that, which keeps the map fair across dialects that differ in how, or
whether, they return path bindings. A disagreement at this level is a genuine
disagreement about answers, not about presentation.

Some cases are constructed so that two different readings of the \emph{same} query text
can be told apart. On a directed two-cycle $m \to n \to m$, the pattern
\verb|{1,4}| yields four paths under \walk{} and two under \trail{}; on a single edge
$u \to v$, an any-direction pattern of length exactly two yields the one path
$u, e_1, v, e_1, u$ under \walk{} and nothing under \trail{}. openCypher has no
restrictor keyword, so its rendering is the same text in both cases; the pair of cases
asks which of the two meanings that one query has.

\section{The map}
\label{sec:map}

\begin{figure}[H]
\centering
\includegraphics[width=\textwidth]{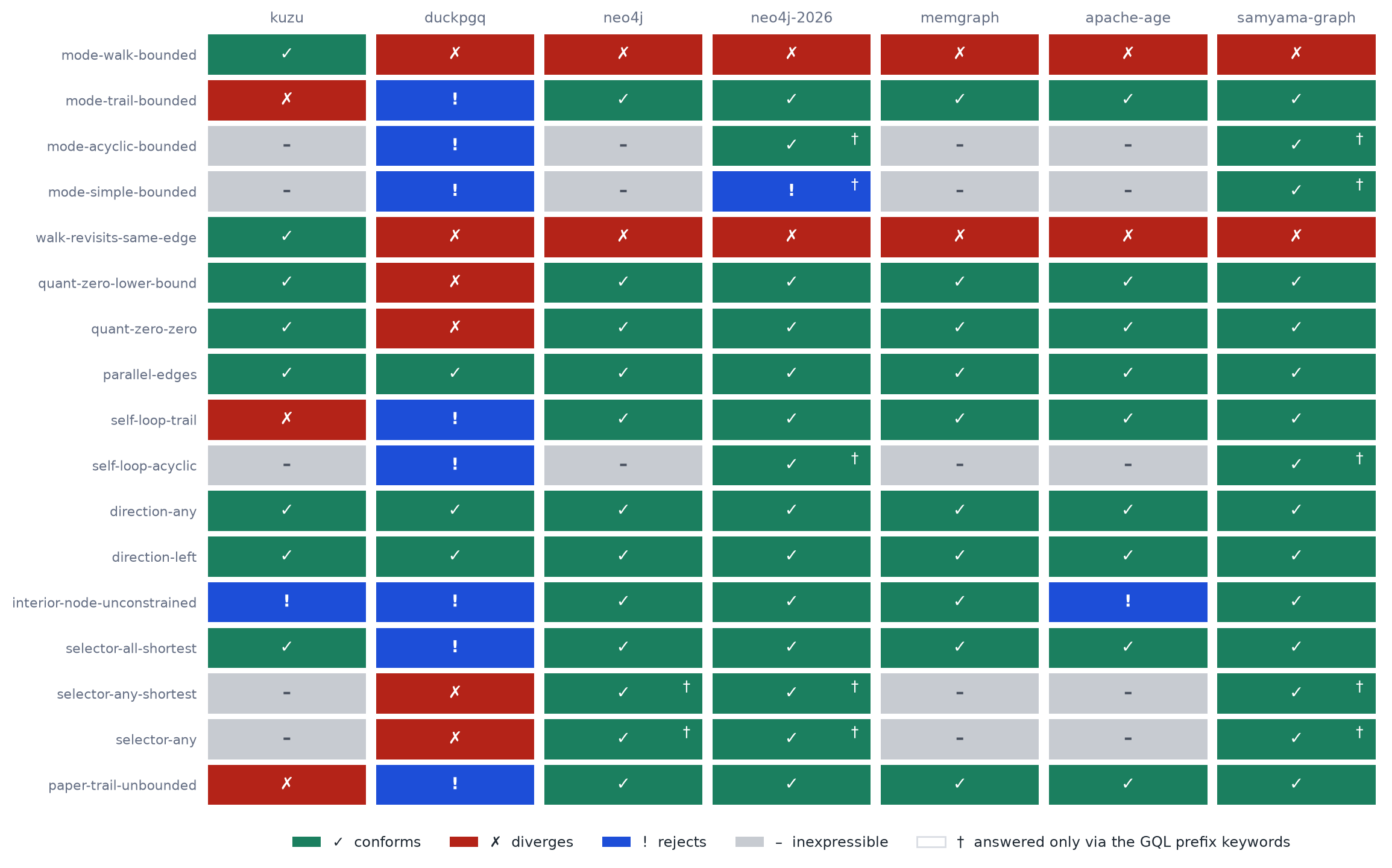}
\caption{The divergence map. Rows are constructs, columns engines. Each cell also
carries a glyph, so the figure survives greyscale printing and colour-vision
deficiency. Every cell is an exact multiset comparison, run three times to confirm the
engine is deterministic; no engine was nondeterministic. The last column is the
authors' own engine, shown for completeness and excluded from every statistic in
Table~\ref{tab:s1s2} and Section~\ref{sec:attribution}.}
\label{fig:map}
\end{figure}

\begin{table}[H]
\centering
\begin{tabular}{@{}lrrrrrr@{}}
\toprule
engine & conforms & diverges & rejects & inexpressible & $S_1$ & $S_2$ \\
\midrule
K\`uzu 0.11.3        & 8  & 3 & 1 & 5 & 0.27 & 0.75 \\
DuckPGQ (DuckDB 1.4.1) & 3  & 6 & 8 & 0 & 0.67 & 0.43 \\
Neo4j 5.26.30        & 12 & 2 & 0 & 3 & 0.14 & 1.00 \\
Neo4j 2026.04.0      & 14 & 2 & 1 & 0 & 0.12 & 0.67 \\
Memgraph 5.9.0       & 10 & 2 & 0 & 5 & 0.17 & 1.00 \\
Apache AGE 1.8.0     & 9  & 2 & 1 & 5 & 0.18 & 0.67 \\
\midrule
headline$^{\ast}$   & 44 & 15 & 11 & 15 & \textbf{0.25} & \textbf{0.58} \\
\midrule
\itshape Samyama-Graph v1.8.0$^{\ddagger}$ (ours) & \itshape 15 & \itshape 2 &
\itshape 0 & \itshape 0 & \itshape 0.12 & \itshape 1.00 \\
\bottomrule
\end{tabular}
\caption{$S_1$ is divergences over the cells the engine answered. $S_2$, the
\emph{silence ratio}, is divergences over divergences plus rejections: the share of
disagreements the user is not told about. The authors' own engine is listed below the
rule and is excluded from the totals and from every other statistic in this paper.
$^{\ast}$ the headline aggregate takes \textbf{one row per product}, at its newest
measured version, so that measuring Neo4j twice does not weight that vendor twice; the
5.26 row is shown but not counted. Over all six external rows the figures are
$S_1 = 0.23$, $S_2 = 0.61$, and over the five products with Neo4j at 5.26 they are
$S_1 = 0.26$, $S_2 = 0.60$. The count of silent divergences is 15 under every one of
the three; only the denominator moves.
$^{\ddagger}$ the authors' own engine, measured at its published release like every
other row, and excluded from the totals. \textbf{Its release carries fixes this suite
prompted}, which is why it scores as it does; see Section~\ref{sec:limitations} before
reading anything into that row. Five of its 15 conforming cells are answered only
through the standard's prefix keywords, which four of the six external releases do not
parse at all; for Neo4j 2026.04 that count is 4 of 14.}
\label{tab:s1s2}
\end{table}

Three readings.

\paragraph{Nine of seventeen constructs are not portable.} Counting distinct answers
per construct over the engines that accepted the query, seven constructs draw exactly
one answer and nine draw two or three. The three-way split is
\verb|mode-walk-bounded|: K\`uzu returns the four walks, every Neo4j release plus
Memgraph and AGE return the two trails, and DuckPGQ returns one row.

\paragraph{The silence ratio is 0.58.} Of 26 disagreements, 15 are silent. A rejection
is a portability problem the user sees at once; a divergence is a portability problem
that ships. Neo4j 5.26 and Memgraph reject nothing in the suite, so every one of their
disagreements is silent --- $S_2 = 1.00$. This is the number we think matters most and
it is not reported anywhere else. It is also the most stable thing in the table: it sits
between 0.58 and 0.61 under all three aggregations, and the numerator --- fifteen silent
divergences --- is identical in every one.

\paragraph{Three products reproduce the standard's own worked example.} On the
\trail{} query whose three bindings the reference
exposition prints, both Neo4j releases, Memgraph and Apache AGE return exactly three. That is
a useful positive control on the whole apparatus, independent of our oracle.

\paragraph{Three engines take the standard's keywords, and no two spell them the same
way.} The suite asks each engine at startup which spelling of the restrictor and selector
prefixes it parses, rather than assuming (Section~\ref{sec:suite}). K\`uzu, DuckPGQ,
Memgraph and Apache AGE take neither. The three that take anything disagree with each
other:

\begin{table}[H]
\centering
\begin{tabular}{@{}llll@{}}
\toprule
& restrictor & selector & \texttt{RESTRICTOR p = \ldots} \\
\midrule
Neo4j 5.26.30   & neither spelling & both spellings & --- \\
Neo4j 2026.04.0 & quantified only  & both spellings & \textbf{rejected} \\
\itshape Samyama-Graph v1.8.0 (ours) & \itshape legacy only & \itshape legacy only &
\itshape \textbf{accepted} \\
\bottomrule
\end{tabular}
\caption{Which spelling of the standard's path prefixes each engine parses, for the
engines that parse any. ``Legacy'' is \texttt{-[:E*1..3]->}; ``quantified'' is
\texttt{(()-[:E]->())\{1,3\}}. ``---'': that release takes no restrictor, so there is
nothing to place before the variable. Neo4j 2026.04 refuses the restrictors in the
legacy form and says so, naming the legacy quantifier in the error. Three engines, three
different subsets, and no overlap in the last column.}
\label{tab:syntax}
\end{table}

The last column is the sharp one. ISO/IEC 39075 heads a path pattern with the restrictor
and binds the variable to the whole thing --- \verb|MATCH TRAIL p = (...)| --- and the
worked examples in the reference exposition are written that way. \textbf{No external
engine we measured parses it.} The single engine that does is the authors' own, and it
does so because this suite reported the gap and its maintainers, who are the authors,
fixed it (Section~\ref{sec:limitations}). That is worth exactly one conclusion and no
more: the order is implementable, so its absence elsewhere is a choice rather than an
obstacle.

\paragraph{The one version pair we have suggests slow convergence, not none.} Neo4j
5.26.30 and 2026.04.0 agree on 14 of 17 constructs. The three that move all go the same
way: \acyc{} becomes expressible and correct on two cells, and \simplem{} is attempted
and rejected. Nothing regressed, and nothing else changed. One vendor pair is not a
trend, but it is the only version-over-version evidence in this paper and it points at
adoption proceeding construct by construct rather than in one release.

\section{Attribution: language difference or implementation defect?}
\label{sec:attribution}

A divergence from the ISO reference is not by itself a defect. openCypher specifies
that a variable-length relationship pattern uses relationship uniqueness --- no
relationship traversed twice within one \verb|MATCH| --- which is \trail{}. An
openCypher engine that returns the trails is doing exactly what its language says. So
we score every answer a second time, against the path mode the engine's own manual
declares: \trail{} for Neo4j, Memgraph and Apache AGE; \walk{} for K\`uzu, which
documents its deviation from openCypher explicitly; and, for DuckPGQ, nothing, because
the dialect it implements \emph{is} the standard.

\begin{figure}[H]
\centering
\includegraphics[width=0.86\textwidth]{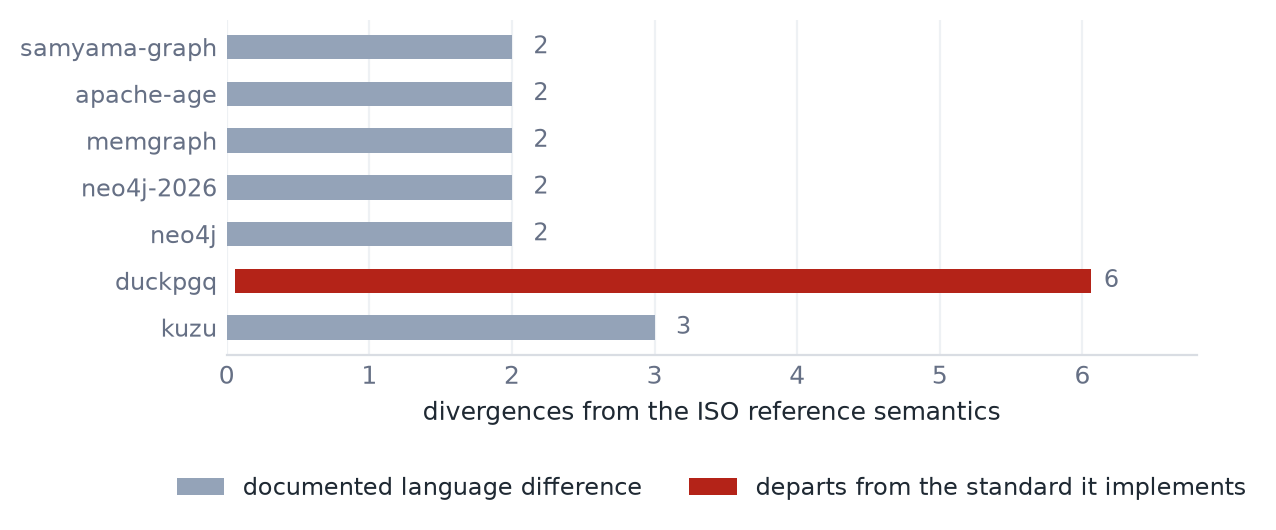}
\caption{Divergences from the ISO reference, split by whether the engine's own
documentation accounts for them.}
\label{fig:attribution}
\end{figure}

Nine of the fifteen divergences are documented language differences. Every K\`uzu and
every openCypher divergence falls in this class: those engines do precisely what they
say they do, and the gap is between the dialect and the standard, not between the
manual and the binary. The remaining six are DuckPGQ's, and DuckPGQ has no second
reading to appeal to.

This split is the practical payload. For a user porting a query, a documented language
difference is a fact to design around; a departure from the implemented standard is a
bug to report. Aggregating the two --- as a bare conformance scoreboard would --- tells
neither audience what to do.

\section{A layer that needs no reference}
\label{sec:meta}

The map is only as persuasive as our reading of the standard, and where a dialect
deliberately deviates it says nothing about whether the engine is \emph{correct}. So we
add a second, weaker but unarguable layer: metamorphic
relations~\citep{chen1998metamorphic,segura2016metamorphic} that hold under
\emph{every} path mode in the standard, because no restrictor mentions the quantifier
bounds and every path has exactly one length.

For $A(lo,hi)$ the answer multiset of a quantified pattern:
\begin{align*}
\text{M1}\quad & A(lo, hi) \subseteq A(lo, hi+1) \\
\text{M2}\quad & A(lo, hi) \subseteq A(lo-1, hi) \\
\text{M3}\quad & A(lo, hi) = \textstyle\biguplus_{k=lo}^{hi} A(k,k)
\end{align*}
An engine that breaks one of these contradicts itself, whatever it believes a path is.

\begin{figure}[H]
\centering
\includegraphics[width=0.86\textwidth]{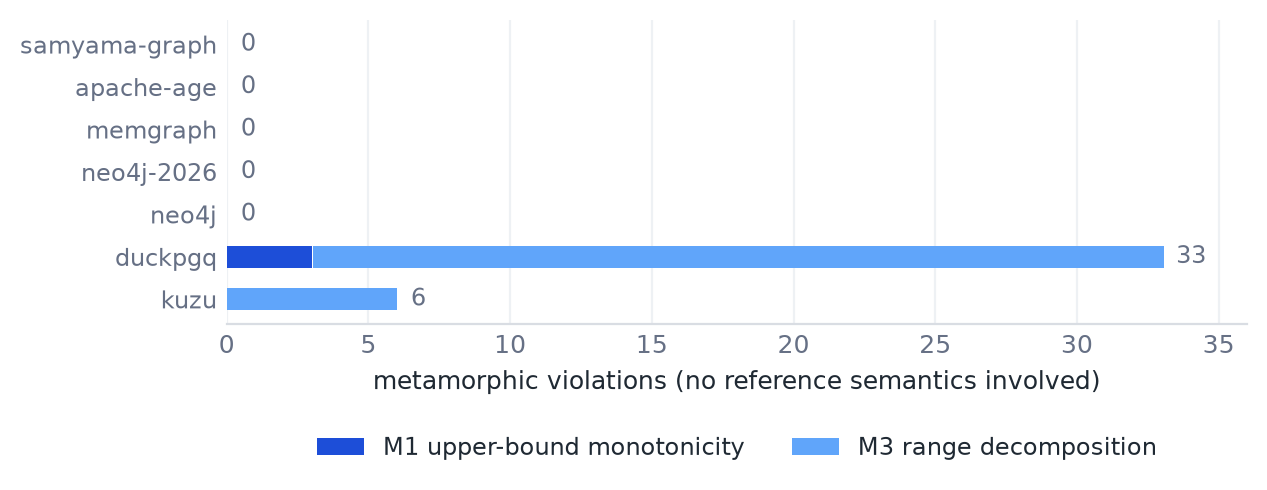}
\caption{Metamorphic violations over six graphs and all quantifier pairs with
$hi \le 4$. Both Neo4j releases, Memgraph, Apache AGE and the authors' own release are
clean; the authors' row enters no count. No query errored on any engine in this run, so
these counts are complete rather than lower bounds.}
\label{fig:meta}
\end{figure}

Thirty-nine violations, all in two engines. They reduce to two defects.

\paragraph{K\`uzu 0.11.3: a lower bound of zero double-counts returns to the start.}
On a graph with a self-loop at $x$ and a two-cycle $x \to y \to x$, the engine reports
\verb|*0..0| $= \{x\}$, \verb|*1..1| $= \{x, y\}$ and \verb|*2..2| $= \{x, x, y\}$, but
\verb|*0..2| $= \{x \times 6,\ y \times 2\}$ where the sum of the three is
$\{x \times 4,\ y \times 2\}$. Every path whose end node is the start node is returned
twice, and the zero-length path is dropped, whenever the lower bound is zero.

\paragraph{DuckPGQ: the quantified pattern is not the standard's, in three ways.}
On a directed two-cycle $m \to n \to m$: \verb|{0,0}| returns a one-hop answer rather
than the zero-length path; \verb|{2,2}| returns nothing although $m, e_1, n, e_2, m$ is
a length-two path --- and the engine's own \verb|{0,2}| does find $m$ at distance two;
and a quantified pattern collapses the multiset of paths to one row per reachable
endpoint, while the unquantified single-edge pattern keeps the multiset. The second of
these is self-contradictory and needs no appeal to the standard at all.

Both are reproduced by standalone scripts in the artifact, each under sixty lines with
no dependency on the rest of the suite.

\section{Limitations and honest negatives}
\label{sec:limitations}

\paragraph{The suite is narrow by design.} Seventeen constructs over the path core.
Writes, transactions, schema, aggregation, subqueries and graph construction are out of
scope, as is performance: this artifact contains no timing number and makes no claim
about speed.

\paragraph{Two engines are not GQL engines and are not scored as failing one.}
Neo4j, Memgraph, Apache AGE and K\`uzu implement openCypher and its descendants, not
GQL. The map places them against the ISO semantics because that is the only common
yardstick, and Section~\ref{sec:attribution} exists precisely so that this placement is
not mistaken for a conformance verdict. Neo4j is measured at two releases, 5.26.30 and
2026.04.0, so that dialect-version drift is visible rather than folded into a single
product row (Section~\ref{sec:map}).

\paragraph{Comparison is at endpoint level.} We compare multisets of endpoint pairs,
not of full path bindings, so that dialects which cannot return path objects are still
measurable. This can only \emph{under}-count divergence: two engines agreeing on
endpoints might still disagree on which paths produced them. Every divergence we report
is therefore real; some divergences are certainly missed.

\paragraph{A query that errors cannot violate a relation.} The metamorphic layer skips
a relation when any query it needs fails rather than counting the failure, so a crash can
hide a violation but never add one. The artifact records every such refusal, and in this
run \textbf{no query errored on any engine}: the counts above are complete, not lower
bounds. On a run where that is not true the artifact says so per engine.

\paragraph{Our reconstruction of the standard's Figure~1 could in principle be wrong.}
It is validated by six independent published answers, which is strong but not a proof.
The reconstruction and its witnesses are published with the artifact so the check can be
repeated.

\paragraph{We did not find a divergence everywhere we looked.} Seven constructs draw a
single answer across every engine that accepted them, including
parallel edges, both direction forms, and \textsc{all shortest}. Those are portable
today, and saying so is part of the result.

\paragraph{Conflict of interest, and how to read our row.} This work was produced at
Samyama, which develops one of the engines measured. It is measured the way every other
engine is: at its published release, v1.8.0. Its row sits below the rule in
Table~\ref{tab:s1s2} and enters no total, no aggregate, and no figure count. No engine's
output, ours included, was used to derive any expected answer; every expected answer
comes from the reference semantics of Section~\ref{sec:ref}, which never consults a
database.

\textbf{That release carries fixes this suite prompted.} An earlier release scored 8 of
17 conforming with 22 metamorphic violations. The suite's findings went to that engine's
maintainers --- who are the authors --- as issues with reproducers, and the fixes shipped:
a quantifier with lower bound zero that collapsed the multiset of paths to one row per
reachable end node, the path restrictors and selectors, the restrictor-then-variable
order, and unbounded memory growth on a graph of two nodes and three edges, which had
made an ordinary read query a denial of service.\footnote{\url{https://github.com/samyama-ai/samyama-graph/issues/1140},
\href{https://github.com/samyama-ai/samyama-graph/issues/1141}{1141},
\href{https://github.com/samyama-ai/samyama-graph/issues/1148}{1148},
\href{https://github.com/samyama-ai/samyama-graph/issues/1183}{1183}.}

\textbf{So the row is not evidence that this engine is better.} No other engine's
maintainers saw these results before their measured release shipped; ours did, and acted
on them within days. What the row is evidence of is narrower and, we think, more useful:
every defect this suite reported was a real defect, and each was fixable in an ordinary
engineering cycle rather than requiring a redesign. The same is available to every vendor
here --- the suite, the reproducers and the issues are public.

The one axis on which our release is not flattered is the one this paper argues matters
most. Its silence ratio is 1.00: it rejects nothing in the suite, so both of its
divergences are silent, and a user meets them as wrong answers rather than as errors.
Neo4j 2026.04 refuses a construct out loud where we simply answer differently, and on
that axis it is ahead of us.

\section{Related work}

\paragraph{GQL and SQL/PGQ theory.} \citet{deutsch2022gpml} give the reference
exposition of GPML; \citet{francis2023digest} the formal semantics of GQL;
\citet{gheerbrant2025expressive} expressive-power results for both languages; and
\citet{thimmaiah2026mgql} a mechanised small-step semantics with a soundness proof for
a read-only fragment. All four verify against the specification. None runs an engine,
which is the gap this paper fills. \citet{thimmaiah2026mgql} in particular is
complementary: a mechanised semantics is a stronger artifact than ours, and connecting
it to the suite is the natural next step.

\paragraph{Path semantics.} Restricting paths to simple paths or trails to keep
answers finite goes back to \citet{mendelzon1995finding}; the complexity landscape for
trail queries is mapped by \citet{martens2020trichotomy}. That literature explains why
the standard offers restrictors at all; it does not say what implementations do.

\paragraph{DBMS testing.} The oracle problem for database engines is attacked by
pivoted query synthesis, non-optimising reference engine construction and query
partitioning~\citep{rigger2020pqs,rigger2020norec,rigger2020tlp}, and guided by plan
coverage~\citep{ba2023qpg}. Those oracles are \emph{internal}: they compare a system to
itself under a transformation, which finds bugs but cannot adjudicate between two
systems that disagree. Our reference semantics is external and adjudicates; our
metamorphic layer is internal in exactly the classical sense and is included because it
survives disagreement about the standard.

\paragraph{Cross-engine measurement.} \citet{rivera2026path} compare engines on path
queries for performance, normalising to trail semantics. \citet{ldbc2015snb} and the
LDBC suites measure throughput on fixed workloads. Neither measures whether the answers
agree.

\section{Conclusion}

Two international standards now define what a graph path pattern means, and the
definitions are precise. We made those definitions executable, gated the result against
the standard's own published answers, and ran it against six releases of five products.

The same pattern draws more than one answer for nine of seventeen constructs, and
more than half of the disagreements are silent. Splitting divergence by whether the
engine's own documentation explains it turns a scoreboard into two actionable lists: a
portability surface for users, and six concrete defects for one engine's maintainers. A
reference-free metamorphic layer adds two more defects, one of them a case where an
engine's answer to \verb|{2,2}| contradicts its own answer to \verb|{0,2}|.

The suite is small enough to read in an afternoon and is open source. We would like it
to become the thing an engine runs before it claims to implement GQL.

\paragraph{Artifact.} Reference semantics, suite, adapters, reproducers, raw results
and figure code are at \url{https://github.com/samyama-ai/gpml-conformance}, under
Apache-2.0. \verb|./run.sh| starts the engines in Docker and regenerates every number
and figure in this paper.

\bibliographystyle{plainnat}
\bibliography{paper23_gpml_conformance}

\end{document}